\documentstyle[preprint,prb,aps]{revtex}
\begin{document}
\draft
\title{Introducing Directionality to Anderson Localization: \\
the transport properties of Quantum Railroads}

\author{C. Barnes$^{1,2}$  B. L Johnson$^2$  G. Kirczenow$^2$}

\address{$^1$ Cavendish Laboratory, Madingley Road,
Cambridge, CB3 0HE, UK}

\address{$^2$ Department of Physics, Simon Fraser University,
Burnaby, British Columbia, Canada V5A 1S6}
\maketitle

\begin{abstract}
We present a study of the transport properties of
a general class of quantum mechanical waveguides:  Quantum Railroads
(QRR) \cite{barnes1}.
These waveguides are characterised by having a different number of
adiabatic modes which carry current in one direction along the waveguide
to the opposite direction; for example $N$ forward modes and $M$
reverse modes.
Just as
Anderson localization is a characteristic of the disordered $N=M$ case and
the quantum Hall effect a characteristic of the $M=0$ case we find that
a
mixed effect,
{\it directed } localization, is a characteristic of QRR's.
We find that in any QRR there are always $|N-M|$ perfectly transmitted
effective adiabatic modes with the remainder being subject to
multiple scattering and interference effects characteristic of the
$N=M$ case.
\end{abstract}

\narrowtext

\section{introduction}

It was Anderson \cite{ander} who first proposed
the idea that there would be `Absence of diffusion in certain random lattices'.
This was the initial insight which gave rise to a
broad class of problems now referred to collectively as \lq Anderson
Localization' \cite{kram}.  The concept of a metal insulator transition
(MIT) in disordered systems
was introduced by
Mott \cite{mottbook,mott}
and subsequently Kunz and Souillard \cite{kunz}, MacKinnon
\cite{mac80},
Czycholl {\it et al} \cite{czy}
and Thouless and Kirkpatrick \cite{thou81} showed in a variety of
different ways that the conductance of a one dimensional system
should vanish with increasing length due to multiple scattering and
interference effects.  The conductance does not vanish in a simple
manner but is characterised by
strong fluctuations.
In weakly
disordered systems the fluctuations have the universal feature that
root mean square conductance is approximately a constant
$ \sigma_g \sim e^2/h $.
This has been seen experimentally
by Blonder, Dynes and White \cite{blond}, Umbach {\it et al}
\cite{umbach} and Webb {\it et al} \cite{webb} and shown
theoretically by
Al'tshuler \cite{altsh} and Lee, Stone and Fukuyama \cite{lee}.
Under more general conditions of disorder, away from the metallic limit,
recent theoretical investigations by
Pendry,
MacKinnon and Pretre \cite{pend}, and MacKinnon \cite{mack} suggest
that here the fluctuations are consistent with an occasional channel being
open when compared to the mean exponential decay. This is the so called
\lq maximal fluctuation theorem'.

An implicit assumption in all of the theoretical work
done on one-dimensional Anderson Localization is that scattering of
waves in these systems occurs between
equal numbers of modes which carry current in both the forward and
reverse directions.
Quasi-one-dimensional systems need not necessarily have this property
however
as has become increasingly clear following the discovery of the quantum
Hall effect by von Klitzing, Dorda and
Pepper \cite{klitz}.
Laughlin \cite{laugh} used a gauge invariance argument to explain
the observed
quantisation of the Hall conductance in a 2D electron gas
to integer multiples of $e^2/h$.
Subsequently, however, following the introduction of magnetic edge
states by Halperin \cite{halper},
Streda, Kucera and
MacDonald \cite{streda}, Jain and Kivelson \cite{jain}, and B\"uttiker
\cite{butt} proposed an alternate
point of view, in which the quantum Hall effect is explained on the
basis of the Landauer \cite{landau} theory of one-dimensional transport.
In these theories edge states, which are the extensions of the quantised
Landau levels at the edges of the system, form the one-dimensional
adiabatic transport channels (or modes) between which disorder motivates
scattering.
This
description of the integer quantum Hall effect in terms of edge states has
gained wide
acceptance.
Although transport in the integer quantum Hall regime is effectively
one-dimensional  Anderson
localization
\cite{ander,kram,mott,altsh,lee,blond,umbach,pend,mack} of the wave
functions does not occur
because {\it all} of the modes on a
given edge of the sample propagate in the {\it same} direction so that
in a macroscopic sample where the edges do not communicate
backscattering of electrons is impossible \cite{butt}.
 Thus
there can be {\it perfect} transmission of electrons through a disordered
macroscopic sample in the quantum Hall regime, in stark contrast to the
zero average transmission found in more conventional 1D systems.

The aim of this paper is to report on a theoretical study of the
general problem of quasi-one-dimensional transport in a disordered
waveguide where scattering occurs between {\it arbitrary}
numbers $N$ and $M$ of adiabatic
modes which carry current in the forward and reverse directions.
We have coined the term
\lq\lq quantum railroad\rq\rq (QRR) \cite{barnes1} as a general term for such
waveguides since a very useful aid in understanding their transport
properties is to think in terms of an imaginary railroad connecting two
cities for example Los Angeles (LA) and New York (NY).  The railroad
consists of $N$ tracks which carry trains in the NY direction and $M$
tracks which carry trains in the LA direction.
Along the tracks there are a number of different switching events,
which represent disorder, and are set
to switch a train from one track
to another either changing its direction of travel or not depending on the
tracks between which it switches.
In studying the transport properties
of QRR's we are essentially asking questions about the likelihood
that a train starting from one city will arrive at the other.
There are two well understood special cases.
The first is  $N=M$ for which, when sufficiently strong disorder is
introduced, Anderson localization of the wavefunctions will occur.  In
the railroad analogy this corresponds to all track forming closed
loops or loops which return to the city they start from so that no
trains pass from one city to the other in either direction.
The other special case is $N \ne 0, M=0$ for which irrespective of the
strength of the disorder we see the quantum Hall effect.  This
corresponds to all track passing in the same direction so that any
switching between tracks does not alter the number of effective
extended rails and hence the ability of trains to pass in the direction
of the N tracks.
For the general case
$N \ne M$ we will show that disorder causes a QRR to exhibit a novel behaviour
that we
will refer to as \lq\lq directed localization\rq\rq \cite{barnes1} in
which the system appears to be in the Anderson localization regime as
far as trains attempting to pass in one direction are concerned but in
the quantum Hall regime for trains passing in the opposite direction.
This corresponds to all the LA tracks and $M$ of the NY tracks forming
closed loops similar to the Anderson case and $N-M$ tracks remaining
extended as in the quantum Hall effect.

To our knowledge no example of a QRR has been reported
experimentally however
both the Azbel-Wannier-Hofstadter (AWH) system
\cite{azbel,wan,hoff}
and two dimensional arrays of
coupled quantum dots
\cite{kir,johns1,johns2,reed,ismail} have been shown to exhibit rich spectra
containing different combinations of rotating ( forward ) and counter rotating
(
reverse ) edge states.
The transmittance between two reservoirs of a Hall bar with an imprinted
AWH system has
been predicted within the edge state picture of
Ramal {\it et al} \cite{ramal3} and MacDonald \cite{macd3} to be
the algebraic sum of
edge states: $|N-M|$.
We have also shown this to be the case for arrays of quantum
dots \cite{barnesqda,akisida}. These systems are therefore
potential realizations of the general QRR.

In section II we will introduce the underlying physics of QRR's by
expanding on the simple railroad concept.
In section III we give a derivation of the limiting
transmittance of a QRR and show how transmission through a QRR is
characterised by directed localization.  In section IV we give the results of a
numerical investigation of the statistical properties of a number of
different QRR's.  In section V we give a derivation for the Shubnikov
de Haas and Hall measurements for a four terminal Hall bar in which each
terminal is connected by a QRR.  We do this both for the case where
disorder is present and where it is not.
Section VI is a summary of our results.

\section{The Simple Railroad}

The picture of a railroad linking NY and LA with trains being
switched between tracks is a remarkably useful model for
understanding the underlying physics of QRR's because it gives a
physical picture for the coexistence of
extended and localized states and indicates what to expect for the
distributions of
quantum mechanical fluctuations.

First consider the simplest example of a railroad
which when disorder is introduced has both extended
and localized track.  This is the case where there
is one LA track (1) and two NY tracks (2,3) shown in
figure~(\ref{six}) event `e'.  In a QRR disorder
promotes scattering between the adiabatic modes so here
we represent it by connecting pairs of tracks.
The six different ways of doing this are shown
shown in
figure~(\ref{six}).   A typical disordered railroad will contain
a random selection of these switching events.
Figure~(\ref{rail21}) shows a typical section of railroad containing seven
switching events $s_3,s_1,s_2,s_3,s_4,s_4,g$.  As far as trains in this
system are concerned it consists of a
single extended NY rail and three closed
loops reminiscent of the $N=M=1$ Anderson case.  The
picture this figure gives is
not an exception but in fact defines a rule.
If we join any pair of switching events together as far as incident
trains are concerned the resultant scattering is equivalent to one of the other
six switching events.
Table~(\ref{multable}) shows the resultant scattering from joining all pairs.
It is clear from this table that the events $s_1,...s_4$ dominate once
introduced into any railroad.  Any railroad with even a single event
from $s_1,...s_4$ will be equivalent to one of these events and
therefore have a single extended state and a set of closed loops and
loops which return to the city they started from.  For example
as far as incident trains are concerned the railroad in
figure~(\ref{rail21}) is equivalent to the event $s_1$.

This picture readily generalises to the case of $N$ NY tracks and $M$
LA tracks.  Here the dominant set of events are those which have
$N-M$ extended NY tracks and no extended LA tracks denoted by $H_M$.
The subscript refers to the number of incident LA/NY tracks which are
reflected into NY/LA tracks by the event.  The symmetry comes about
because for each NY track that is reflected it takes the space of an LA
track thereby forcing and LA track to also be reflected.  When we join
an event from $H_M$ to any other event
the result must be in $H_M$ since we cannot increase the number of
extended tracks which pass across any scattering event.
Hence we expect that a typical railroad
will consist of a set of $N-M$ extended NY tracks and that the
remaining $M$ NY tracks and $M$ LA tracks will form closed loops
similar in form to the $N=M$ case. The algebraic structure of the
general case is discussed in the appendix.

For the quantum mechanical model we will show that
we have a coexistence of extended and
localised states similar to the simple railroad.
In this model however when the localised states are
closely degenerate or coupled to reservoirs we
would expect to see random fluctuations due to tunnelling
appearing above the simple railroad predictions
for the transmittances.

As we have said for the general case disorder causes closed loops
similar to the $N=M$ case
the difference being that the $N-M$ extended rails fill space and
therefore on average force the closed loops apart.  Thus we might
expect that the average transmittance
would decrease with increasing $N-M$
but that the form of the distribution and the variance be dependent
only upon $M$.  This discussion will be the subject of our section on
numerical work.

\section{The Quantum Railroad}

A quantum mechanical picture of the railroad with $N$ NY
tracks and $M$ LA tracks
must include a number of features.  We must talk in terms of electron
wavefunction amplitudes;
the tracks become the adiabatic modes of a quantum
mechanical waveguide.
Scattering probabilities must be continuous;
when a mode is scattered it will be scattered with
a different random amplitude into
all available modes.  If we wish to conserve the number of
electrons entering and
leaving the waveguide then each scattering event must be unitary.
Experimentally
for such waveguides we would be interested in making standard
magneto-resistance
measurements such as those which are used for the Hall effect and
Shubnikov de Haas effect \cite{klitz}.
We will discuss this point in more detail in section V.
The mathematical constituents to
such measurements are the two point transmittances and reflectances.
 From them we
can construct the experimentally measured quantities using the
B\"uttiker formalism \cite{butt}.
At this stage we will
switch to the notation
`forward' for the $N$ modes and `reverse' for the $M$ modes we will also
assume that the forward direction is the majority mode direction; $N
\ge
M$.

We will prove that as we
introduce more scatterers into a QRR the transmittance in the forward
direction reduces to $T=N-M$ and in the reverse direction to $T'=0$.
Also, in analogue with the simple railroads
effective open rails and closed loops of track, we show that a QRR has
a set of $N-M$  perfectly transmitting forward channels and a set $2M$
channels for
which transmission coefficients are determined by multiple scattering and
interference effects.
Our proof comes in three parts.
The first is to show that unitarity imposes lower limits on the
values that $T$ and $T'$ may take.  The second part is to
make a justification for
stating that these limits are achieved by increasing the number of scattering
events in the QRR and the third part is to show that the transmission
probabilities
in the diagonal basis defined by the
disorder in a QRR are equal to one for a set of $N-M$ channels and the
remainder are
equally matched between the forward and reverse directions.

First we will prove that there are lower limits imposed on the values of the
forward transmittance $T$ and the reverse transmittance
$T'$.
The scattering matrix of a QRR will have the form
\begin{equation}
{\bf S}= \left( \begin{array}{cc}
{\bf T} & {\bf R} \\
{\bf R}' & {\bf T}'
\end{array} \right)
\end{equation}
where ${\bf T}$ is an $N \times N$ transmission matrix containing the
complex {\it
amplitudes } for scattering between the $N$ forward modes,
${\bf T}'$ is the $M \times M$ transmission matrix for the
$M$ reverse modes and ${\bf R}$ and ${\bf R}'$ are the corresponding
reflection matrices.

The two-terminal transmittances $T,T'$ and reflectances $R,R'$ of the QRR,
when connected at either end to perfectly emitting and absorbing
reservoirs, are simply related to the elements of this ${\bf S}$ matrix through
the norms of the transmission and reflection matrices
\cite{landau2,fisher,stone}
\begin{equation}
T  = ||{\bf T}||^2 \\
T' = ||{\bf T}'||^2 \\
R  = ||{\bf R}||^2 \\
R' = ||{\bf R}'||^2
\label{trandefs}
\end{equation}
In this case the norm is defined by the inner product
\begin{equation}
({\bf a},{\bf b}) = {\it trace}({\bf a}{\bf b}^{\dag} )
\label{proddefs}
\end{equation}
where ${\bf a}$ and ${\bf b}$ are matrices with suitable dimensions.

 From these definitions it is easy to see that in a QRR containing no
scattering the transmittances will have the form
$T=N$ in the forward direction and $T'=M$ in the
reverse direction since each mode carries the same current.
For such a system where no scattering occurs it has been shown that a
fractional quantum Hall effect will be observed \cite{johns1,johns2}.
We will discuss this
point in section V.

If we now introduce a series of unitary scatterers into the QRR
then
the scattering matrix for the whole system must also be unitary.
This
condition implies the following set of relations between the reflection
and transmission matrices
\begin{eqnarray}
{\bf T }{\bf T }^{\dag} + {\bf R }{\bf R }^{\dag} & = & {\bf 1}_N
\label{unitdefs1} \\
{\bf T'}{\bf T'}^{\dag} + {\bf R'}{\bf R'}^{\dag} & = & {\bf 1}_M
\label{unitdefs2} \\
{\bf T }^{\dag}{\bf T } + {\bf R'}^{\dag}{\bf R'} & = & {\bf 1}_N
\label{unitdefs3} \\
{\bf T'}^{\dag}{\bf T'} + {\bf R }^{\dag}{\bf R } & = & {\bf 1}_M
\label{unitdefs4}
\end{eqnarray}
Taking the norms of these relations we find that they imply the
following set of relations between
the reflectances and transmittances
\begin{equation}
T + R = N \qquad
R = R' \qquad
T'= T - (N-M)
\label{relats}
\end{equation}
These three relations together with the fact that $T,T',R,R'$ are real
and positive imply that any such system will have transmittances in the
ranges
\begin{eqnarray}
0 & \le & T' \le M \\
N-M & \le & T \le N
\end{eqnarray}
Hence, we see that the transmittance in the majority mode direction has
a lower limit of
 $T=N-M$
and in the minority mode direction $T'=0$.

The second part of our proof is to make a justification for asserting
that as we add more scattering events to our QRR
both $T$ and $T'$ will tend to decrease
and therefore for a typical system will eventually reach their
minimum values.
Adding a unitary scatterer to a QRR
will cause a change in its reflection matrix given by
\begin{equation}
{\bf R}_{+1} = {\bf R} +{\bf B}
\label{newref}
\end{equation}
where
\begin{equation}
{\bf B}= {\bf T r}({\bf 1} - {\bf R' r} )^{-1} {\bf T'}
\end{equation}
Capitals represent the transmission and reflection matrices of the
initial QRR and lower case those of the added scatterer.
Taking the norm of Eq.~(\ref{newref}) and using Eqs.~(\ref{relats})
we
find
\begin{equation}
T_{+1} = T - ||{\bf B}||^2 - ({\bf B,R}) - ({\bf R,B})
\label{lower}
\end{equation}
Note that the same relation holds for $T'$.
Since the last two terms in Eq.~(\ref{lower}) may be positive or negative
it is clear that adding the extra scatterer can cause
the transmittance to increase or decrease.  However the possible choices
for making $T_{+1}$ larger are limited in comparison to those which make
it smaller and also the range of those choices is highly dependent on the
details of the ${\bf S}$ matrix of the initial QRR.  Thus unless the
the scatterers are added to the QRR with
properties
highly correlated to what has come before the transmittance will tend to
decrease.
For example, if we construct a QRR by adding uncorrelated
random unitary scatterers, because there are
no correlations between the reflection phases of such scatterers it is
easy to show that
$\overline{({\bf R,B})}=0$
and $\overline{({\bf B,R})}=0$.  The bar indicates an
average over all possible scatterers which may be added.
The proof of this is simple.
Both
inner products may be expanded as
multinomial series in the reflection amplitudes contained in ${\bf r}$
\cite{crispproc}.
After averaging, these series become
expansions in the multivariate moments of the elements of ${\bf r}$.  Each of
these moments is zero if there are no correlations between the
reflection phases and therefore on average the inner
products are zero. This fact indicates that on average adding
scatterers to the QRR
will reduce its transmittance in either direction.
This type of random phase model has been used many times in the investigation
of
Anderson localization and in particular in connection with its use in
determining a
scaling theory\cite{scaande1,scaande2,scaabra,scaande,scasakj,scaston}.

If adding an extra unitary scatterer to a QRR typically
reduces its transmittance in both directions then for systems
containing many scatterers
experiments will measure the values
\begin{eqnarray}
T' &=& 0  \\
T  &=& N-M
\end{eqnarray}
The possibility that they may reduce to different minima is excluded by
the fact that ${\bf B}={\bf 0}$ typically only if ${\bf T'}={\bf 0}$.
The QRR therefore is a system in which the transmittance is directed. In
one direction the system is opaque and in the other it has a transmittance
which tends to a quantised value with increasing length.

The last part of our proof is to show for any QRR that there are a
set of $N-M$ perfectly transmitted effective channels and that
scattering between the remaining effective channels has the signature
of
the $N=M$ case (
for each forward channel with transmission probability
$|\lambda_i|^2 < 1$ there is a reverse channel with the same transmission
probability )
for which it is known that multiple
scattering and interference effects cause Anderson localization.

So far we have considered the transmission and reflection matrices in
the basis of the adiabatic modes of the undisordered QRR however when
disorder is introduced to the system it defines another unitary basis in which
the transmission matrix is diagonal.  In general the basis may be
different on either side of the QRR so that (note that we only write
the equations for the forward direction but identical expressions hold
for the reverse direction)
\begin{equation}
{\bf P T Q^{\dag}} = {\bf \Lambda }
\end{equation}
Where ${\bf P}$ and {\bf Q} are the matrices containing the bases for
the left and right side respectively and ${\bf \Lambda }$ is a diagonal
matrix containing the transmission amplitudes $\lambda_i$ of the new
basis.
The transmission probabilities for these new channels $|\lambda_i|^2$
are simply the eigenvalues of ${\bf T^{\dag} T}$ since
\begin{equation}
{\bf Q T^{\dag} P^{\dag} P T Q^{\dag} } = {\bf Q T^{\dag} T Q^{\dag} }
= {\bf \Lambda^* \Lambda }
\end{equation}
In order to show that $N-M$ of these transmission probabilities are
equal to unity and the others exactly match between the forward and
reverse directions it is convenient to
look at sums of their $n$th powers.  From
equations~(\ref{unitdefs1},\ref{unitdefs4}) we find
\begin{eqnarray}
trace(({\bf T T^{\dag} })^n ) &=& N + \sum_{i=1}^{n} (-1)^i {n \choose i} ({\bf
R R^{\dag}})^i \label{ttdag} \\
trace(({\bf T'^{\dag} T' })^n ) &=& M + \sum_{i=1}^{n} (-1)^i {n \choose i}
({\bf
R^{\dag} R})^i \label{tptpdag}
\end{eqnarray}
Using the identities
\begin{eqnarray}
trace( ({\bf A A^{\dag} } )^n ) &=& trace( ({\bf A^{\dag} A} )^n ) \\
                           &=& \sum_{i=1}^m |\nu_i|^{2n}
\end{eqnarray}
where ${\bf A}$ is any $m \times m'$ matrix and $\nu_i$ are the
eigenvalues of ${\bf A}$,
these expressions reduce to
\begin{equation}
\sum_{i=1}^N |\lambda_i|^{2n} = \sum_{i=1}^M |\lambda'_i|^{2n} + N-M
\end{equation}
When we take the limit $n \rightarrow \infty$ of this expression all
transmission probabilities for which $|\lambda_i|^{2n} < 1$ reduce to
zero  so that in order to satisfy the identity we must have at least
$N-M$ forward transmission probabilities equal to unity. If the system
is sufficiently long then there will be exactly $N-M$ with unit
transmission probability in the forward direction and the rest will be less
than unity.
In this limit cancelling the unit transmission probabilities from each side of
the expression we are left with
\begin{equation}
\sum_{i=1}^M |\lambda_i|^{2n} = \sum_{i=1}^M |\lambda'_i|^{2n}
\end{equation}
We arrange the transmission probabilities largest to smallest then take the
limit $n
\rightarrow \infty $.  The largest transmission probabilities must be equal
since the other transmission probabilities contribute no weight to
the sums in comparison.
These
then cancel from the sum and we can make the same argument for
successive pairs.

Hence we have shown that transmission through a QRR is carried by $N-M$
perfectly transmitted effective channels and that fluctuations
are caused by the remaining $2M$
effective channels which will have transmittances determined by
multiple scattering and interference effects within the QRR.
Fluctuations in the transmittance can of course play an important role
in determining whether a particular QRR will have a forward
transmittance $T=N-M$ and reverse transmittance $T'=0$.  Localized
states or \lq necklaces' of localized states \cite{necklace} can act like
like pinholes through an otherwise opaque system.  We look more closely
at the statistical properties of QRR's in the next section using
numerical techniques.

\section{Numerical Studies}

For the Anderson case $N=M$ it is well known that the
transmittance from one disordered to sample to another fluctuates
wildly over many orders of magnitude and that these fluctuations
grow with system
size \cite{scaabra,stone82,kirk84a,kirk84b,abri81,melni81}.
The distribution of possible transmittances for any ensemble of
disordered systems does not obey the central limit theorem as
its moments are always dominated by rare but strong fluctuations
\cite{pend,mack,pend2}.
However the distribution of the logarithm of the transmittance is
a more tractable quantity \cite{oconn75} which for the bulk of the probability
is approximately normally distributed \cite{marcos93a}.  The logarithm of the
transmittance relates directly to the exponential decay of wave
functions into a system from a reservoir
or decay away from a localised centre of electron probability in the
interior of a system.  The usual notation is  $T=e^{-\gamma L}$ where
$\gamma$ is the inverse localization length and $L$ is the length of
the system.
Our calculations in the previous section have shown that the transport
in a general QRR is carried by $N-M$ perfectly transmitted channels and
that the remainder of the channels participate in multiple scattering
and interference effects identical in character to the unitary scattering
 of the Anderson case.  In this section we produce numerical simulations which
show this relation explicitly and illuminate the differences
which are
easily explained with the insight gained from the simple railroad model.

Our numerical work is based on a matrix method which consists of
generating a transfer matrix of the form
\begin{equation}
{\bf T}_i= \left( \begin{array}{ccc}
 {\bf t - r{t'}^{-1}r'} &  \quad  & {\bf r {t'}^{-1}} \\
-{\bf {t'}^{-1} r' }    &  \quad  & {\bf {t'}^{-1} }
\end{array} \right)
\label{transfer}
\end{equation}
for each
scattering event $i=1 \rightarrow L$ and multiplying them together
to find the
transfer matrix for the whole system
\begin{equation}
{\bf T}=\prod_{i=1}^{N} {\bf T}_i
\end{equation}
The algebraic expressions in the blocks of the resultant transfer matrix are
the same as in equation~(\ref{transfer}) for the transmission and
reflection matrices of the whole system and therefore are straight
forwardly
rearranged to find the
transmittance and reflectance of the whole system.
We consider a random unitary ensemble of scatterers
because this ensemble contains both
scattering with time reversal symmetry and without and therefore
constitutes a rigourous test for the general case.
Our method for generating the random unitary matrices for each
scattering event
is as follows:
first we
generate a set of $N+M$ random complex vectors which lie in a complex
unit $N+M$ dimensional sphere;
then we
normalise each vector to have unit length and perform Gram-Schmidt
orthogonalisation to find a set of $N+M$
orthogonal vectors e.g $u^{\dag} u =1$.
These vectors then form the column vectors of a unitary matrix which
may be partitioned to give the matrices ${\bf t,t' r,r'}$.
We found that this
method together with a good random number generator gave
an even distribution for
the phase of the determinants of the unitary matrices
 which was our criterion for randomness.
For error checking in this method we always calculate both
the reflectance and
transmittance and check to see that unitarity is preserved to
better than 0.1\% of
the minority direction transmittance.

This numerical method straight forwardly demonstrates how a typical
QRR achieves the condition for directed localization.
Figure~(\ref{decay21}a) shows the $N=2, M=1$ case for seven different
configurations of scattering events.  Each trace shows the logarithm
of the minority mode
transmittance as a function of the number of scattering events.
All traces show that the transmittance reduces to zero very quickly
for these typical
configurations.  The relation~(\ref{relats}) implies that if the minority
mode transmittance
reduces to zero then the majority mode transmittance takes on the value
 $T=N-M$.
Figure~(\ref{decay21}b) shows the $N=M=2$ Anderson case
for a number of different
impurity configurations.  It is
clear here also that the transmittance reduces to zero.

By calculating the transmittances of large numbers of randomly
generated
QRR's we can demonstrate their statistical properties.
Figures~(\ref{meq1}a,b,c)
show a series of histograms of the natural
logarithm of
the minority transmittance $log(T')$ against its probability of occurring
$P(log(T'))$
after 30 random unitary scattering events for different choices of $N$
and $M$.  Each histogram was
generated by calculating the minority mode transmittance of
approximately 32000 different samples.
We list the mean, rms value and ratio of the mean to the rms
value for each distribution in table~(\ref{values}).  These histograms
show in a quantitative manner that a typical experimental sample would
be expected to have a transmittance well quantised to $T=N-M$.  The
worst case $N=M=3$ has less than four percent of it transmittances
greater than $T'=0.03$.
The approximate shape of these distributions is normal as is expected
given our analytic comparison with the Anderson case
\cite{marcos93a}.  We can explain
the other features easily on the basis of the simple railroad model.

First looking at figures~(\ref{meq1}a,b,c) it is clear that
the shape distributions and rms values have only a weak dependence on $N$ but
a strong dependence on $M$.
This is accounted for by the simple railroad picture
in which the fluctuations are caused by tunnelling through localised
states because for the general case the form of localized states is
dependent only upon $M$.
Secondly we note that the distributions
appear to increase in width with decreasing $M$.  This may be accounted for
by the fact that the larger the value of $M$ the easier it is to
get around a particular impurity or be de-localized since there
are more states available to scatter into and therefore the
fluctuations will be weaker.  This fact also accounts for
the the fact that the mean transmittance increases with increasing
$M$.
Finally we note that the mean of
each distribution decreases with increasing $N-M$.  This is
consistent with our observation that for the simple railroad,
although the scattering occurs between $M$ forward modes and $M$
reverse modes, the $N-M$ extended modes effectively force these
states further apart making it harder to pass through the
system.

\section{Hall and Shubnikov de Haas Measurements}

Conditions of electrodynamic continuity prohibit connecting a single
QRR between two electron reservoirs.  With zero voltage difference
between the reservoirs such a device would cause current to pass from
one to the other in the majority mode direction.   QRR's must
be connected between reservoirs such that the total number of quantum
mechanical modes which carry current into and out of each reservoir is
equal.
For example a Hall bar configuration
where the QRR's are composed of edge states.
Hall
and Shubnikov de Haas measurements are made on these systems using four
reservoirs passing current between two and measuring the resulting
voltage between the other two \cite{klitz}.
In this section we will calculate these quantities
for  the case where no scattering is
present and then for the case where scattering is present and
directed localization
occurs.

The case where no scattering occurs leads to fractional Hall and Shubnikov
de Haas conductances as has been discussed for the case of multiple edge state
transport in arrays of quantum dots \cite{johns1,johns2}.
Figure~(\ref{nodis}a) shows the picture we have in mind for this case.
The expression for current passing out of a
reservoir $\alpha$ in a multi-probe system is given by the B\"uttiker
formalism \cite{butt}
\begin{equation}
I_\alpha = {e \over h } \left(  \mu_\alpha ( N_\alpha -R_\alpha) -
\sum_{\beta \ne \alpha} \mu_\beta T_{\beta \alpha} \right)
\end{equation}
so that the four leads in figure~(\ref{nodis}a) we will have
currents
\begin{eqnarray}
{h \over e}I_s &=& \mu_s (N+M) - \mu_1 M - \mu_d N  \\
{h \over e}I_1 &=& \mu_1 (N+M) - \mu_2 M - \mu_s N  \\
{h \over e}I_2 &=& \mu_2 (N+M) - \mu_d M - \mu_1 N  \\
{h \over e}I_d &=& \mu_d (N+M) - \mu_s M - \mu_2 N
\end{eqnarray}
For the Shubnikov
de Haas measurement we pass a current $I$
between probes $\mu_s$ and $\mu_d$ and
measure the voltage between probes $\mu_1$ and $\mu_2$.
No current flows into or out of the voltage probes by definition so that
$I_1 = I_2=0,
I_s = -I_d=I $.
The Shubnikov de Haas resistance is defined by
\begin{equation}
R_{S de Haas} = {{ \mu_1 - \mu_2 } \over eI }
\end{equation}
which solving the above set of equations gives
\begin{equation}
R_{S de Haas} = {h \over e^2}{ NM \over {N^2M + NM^2 +N^3 +M^3}}
\end{equation}
For the Hall measurement we pass a current $I$ between probes $\mu_1$
and $\mu_d$ and measure a voltage between $\mu_s$ and $\mu_2$.  This
gives us $I_s = I_2 =0$ and $ I_1=-I_d =I$.  Using the
definition.
\begin{equation}
R_{Hall} = {{ \mu_s - \mu_2 } \over eI }
\end{equation}
and solving the current relations we find
\begin{equation}
R_{Hall} = {h \over e^2}{ {N-M} \over {N^2 + M^2} }
\end{equation}
As has been noted these equations for the Hall and Shubnikov de
Haas resistances
imply a fractional conductance for certain values of $N$ and $M$.

If we now consider the case where directed localization occurs
then a different picture
appears: figure~(\ref{nodis}b).  The set new of current relations have the form
\begin{eqnarray}
{h \over e}I_s &=& \mu_s (N+M - 2M) - \mu_d (N-M)  \\
{h \over e}I_s &=& \mu_1 (N+M - 2M) - \mu_s (N-M)  \\
{h \over e}I_s &=& \mu_2 (N+M - 2M) - \mu_1 (N-M)  \\
{h \over e}I_s &=& \mu_d (N+M - 2M) - \mu_2 (N-M)
\end{eqnarray}
These equations when simplified are familiar: they are
exactly those of
the ordinary integer edge state picture but with the usual $N$
replaced by $N-M$. They solve to give
\begin{eqnarray}
R_{S de Haas} & = & 0 \\
R_{Hall}      & = & {h \over e^2} {1 \over {N-M}}
\end{eqnarray}

The Hall bar with no scattering and
therefore no directed localization yields fractional
Shubnikov de Haas and Hall
conductances for certain values of $N$ and $M$ and the Hall bar
with scattering and
therefore directed localization
yields a simple Hall and Shubnikov-de Haas effect.

\section{summary}

In summary, we have given a detailed theory for the transport
properties of a general class of disordered waveguides the quantum
railroads (QRR's).  We have shown analytically that the
effect of disorder in a
QRR with $N$ channels which carry current in the forward direction and
$M$ channels which carry current in the reverse direction is to cause
the transmittance to equilibrate to $T=|N-M|$.  We have also shown
analytically that current is carried through a QRR by $|N-M|$ perfectly
transmitted effective channels and that the remaining $2M$ channels
participate in multiple scattering and interference effects identical
in nature to the Anderson case where $N=M$.   Our numerical simulations
confirm this by giving normal distributions for the logarithm of the
minority mode transmittance and
further show that the mean decay length in these systems decreases with
increasing $|N-M|$ consistent with the notion that the $|N-M|$ fully
transmitted states fill space and therefore force any localized states
apart thereby making the transmittance on average smaller.

\section{appendix:algebraic properties}

The set of all scattering events for the general case of $N$ NY tracks
and $M$ LA tracks contains $(N+M)!$ elements.  Intuitively we might
think that this set with the composition rule defined by joining two
events would be related to the symmetric group of order
$N+M$ it is not though.
The under the composition rule the set in fact forms a monoid \cite{mono} for
which the identity element is the event where no scattering occurs.
This monoid naturally
splits into one sub-group and a set of $M$ semi-groups.  The sub-group is the
set $H_0$ which contains all the events with $N$ extended NY tracks and
$M$ extended LA tracks.
The $M$ semi-groups are
the sets $H_i$ with $i=1,...,M$ where $H_i$ is the set of events which
have $N-i$ extended
NY tracks and $M-i$ extended LA tracks.  These semi-groups form
a hierarchy such
that when elements from $H_i$ are composed with elements from $H_j$
if $i \ge j$ the result is an element in $H_i$.  This makes $H_M$
the ideal semi-group of the monoid and it is to this property that we may
ascribe the dominant role of $H_M$ in any railroad.

\acknowledgements

We wish to thank A. H. MacDonald for a stimulating discussion.
C. B. would like to thank the
Royal Society of London for their financial support and SFU
for their kind hospitality during his stay in Canada.
B. L. J. and G. K. wish to
acknowledge the financial support of the NSERC of Canada and the CSS at
SFU. C.B. would like to thank D. Maslov and K. Yeo for useful
discussion.

\begin{figure}
\caption{The six possible scattering events for the $N=2,M=1$ simple
railroad.}
\label{six}
\end{figure}

\begin{figure}
\caption{A typical stretch of railroad for the $N=2,M=1$ case showing a
single extended track and a number of localized tracks.}
\label{rail21}
\end{figure}

\begin{figure}
\caption{A typical stretch of railroad for the $N=4,M=2$ case showing
two extended tracks (a) and a number of localized tracks (b).}
\label{rail42}
\end{figure}

\begin{figure}
\caption{The log of the minority mode transmittance for
a) $N=2,M=1$ and b) $N=M=2$,
as a function of
the number of scatterers in the system each for seven different
impurity configurations.}
\label{decay21}
\end{figure}

\begin{figure}
\caption{Histograms of the probability distributions of the minority
mode transmittances for
a) $M=1$ and $N=1,2,3,4$,
b) $M=2$ and $N=2,3,4,5$,
c) $M=3$ and $N=3,4,5,6$,
calculated for railroads with 30
scattering events.}
\label{meq1}
\end{figure}

\begin{figure}
\caption{a) Edge state transport with no scattering in a Hall bar.
b) Edge state transport with directed localization in a Hall bar.}
\label{nodis}
\end{figure}

\mediumtext
\begin{table}
\caption{ Multiplication table for the $N=2,M=1$ railroad
switching events.}
\label{multable}
\begin{tabular}{|l|llllll|}
\multicolumn{1}{|l|}{$\times$}
&\multicolumn{1}{l}{$e  $}
&\multicolumn{1}{l}{$g  $}
&\multicolumn{1}{l}{$s_1$}
&\multicolumn{1}{l}{$s_2$}
&\multicolumn{1}{l}{$s_3$}
&\multicolumn{1}{l|}{$s_4$}  \\
\tableline
 $e  $ &  $e  $& $g  $ & $s_1$ & $s_2$ & $s_3$ & $s_4$ \\
 $g  $ &  $g  $& $e  $ & $s_4$ & $s_3$ & $s_2$ & $s_1$ \\
 $s_1$ &  $s_1$& $s_2$ & $s_1$ & $s_2$ & $s_2$ & $s_1$ \\
 $s_2$ &  $s_2$& $s_1$ & $s_1$ & $s_2$ & $s_2$ & $s_1$ \\
 $s_3$ &  $s_3$& $s_4$ & $s_4$ & $s_3$ & $s_3$ & $s_4$ \\
 $s_4$ &  $s_4$& $s_3$ & $s_4$ & $s_3$ & $s_3$ & $s_4$ \\
\end{tabular}
\end{table}

\mediumtext
\begin{table}
\caption{Table of the means and rms values of each
histogram in figures~(\protect\ref{meq1}a,b,c).}
\label{values}
\begin{tabular}{|cccc|}
\multicolumn{1}{|c}{N,M}
&\multicolumn{1}{c}{Mean}
&\multicolumn{1}{c}{RMS}
&\multicolumn{1}{c|}{$|$Mean/RMS$|$} \\
\tableline
\tableline
1,1 & -41.54 & 11.66 & 3.56 \\
2,1 & -53.30 & 9.76  & 5.46 \\
3,1 & -61.00 & 8.87  & 6.88 \\
4,1 & -67.06 & 8.30  & 8.08 \\
\tableline
2,2 & -15.80 & 5.54  & 2.85 \\
3,2 & -25.81 & 5.41  & 4.77 \\
4,2 & -33.13 & 5.28  & 6.27 \\
5,2 & -39.00 & 5.12  & 7.61 \\
\tableline
3,3 & -10.50 & 3.99  & 2.63 \\
4,3 & -17.50 & 3.99  & 4.38 \\
5,3 & -23.28 & 3.97  & 5.86 \\
6,3 & -28.09 & 3.91  & 7.18 \\
\end{tabular}
\end{table}

\end{document}